\documentclass[twocolumn,showpacs,preprintnumbers,amsmath,amssymb,prl,superscriptaddress]{revtex4}
\usepackage{amsmath,amssymb,amsthm,color}
\usepackage{graphicx}
\usepackage{dcolumn}% Align table columns on decimal point

\newcommand {\beq}{\begin{eqnarray}}
\newcommand {\eeq}{\end{eqnarray}}

\newcommand {\BiSe}{{Bi$_2$Se$_3$}}

\begin{document}
%%%%%%%%%%%%%%%%%%%%%%%%%%%%%%%%%%%%%%%%%%%%%%%%%%%%%%%%%%%%%%%%%%%%%%%%
%%%%%%%%%%%%%%%%%%%%%%%%%%%%%%%%%%%%%%%%%%%%%%%%%%%%%%%%%%%%%%%%%%%%%%%%
\preprint{CALT-68-2848, IPMU11-0173, NSF-KITP-11-273}

\title{Instability in magnetic materials with dynamical axion field}

\author{Hirosi Ooguri}
\affiliation{California Institute of Technology, 452-48, Pasadena, California 91125, USA}
\affiliation{Kavli IPMU,
University of Tokyo (WPI), Kashiwa 277-8583, Japan}
\author{Masaki Oshikawa}
\affiliation{Institute for Solid State Physics, University of Tokyo,
Kashiwa 277-8581, Japan}

%\date{\today}% It is always \today, today,
             %  but any date may be explicitly specified

%-----------------------------------------
\begin{abstract}
It has been pointed out that the axion electrodynamics exhibits
instability in the presence of a background electric field. 
We show that the instability leads to a complete screening of 
an applied electric field above a certain critical value
and the excess energy is converted into a magnetic field. 
We clarify the physical origin of the screening effect
and discuss its possible experimental realization in 
magnetic materials where magnetic fluctuations
play the role of the dynamical axion field.
\end{abstract}

\pacs{14.80.Va,73.61.-r}
% PACS, the Physics and Astronomy
                             % Classification Scheme.
%\keywords{Suggested keywords}%Use shokeys class option if keyword
                              %display desired
\maketitle

%%%%%%%%%%%%%%%%%%%%%%%%%%%%%%%%%%%%%%%%%%%%%%%%%%%%

\noindent
\textit{Introduction.---}
The electrodynamics is a $U(1)$ gauge theory usually
defined by the Maxwell action with the Lagrangian density
\begin{equation}
 {\cal L}_{\rm EM} =
\frac{1}{8\pi}  (\epsilon \vec{E}^2 - \frac{1}{\mu} \vec{B}^2),
\label{eq.L_EM}
\end{equation}
where $\vec{E}$ and $\vec{B}$ represent electric
and magnetic fields, and the permittivity $\epsilon$
and permeability $\mu$ are both unity in vacuum.
Gauge invariance allows an additional term in the Lagrangian density
\begin{equation}
 {\cal L}_{\theta} = \frac{\alpha}{4\pi^2}
  \theta \vec{E} \cdot \vec{B} ,
\label{eq.Ltheta}
\end{equation}
where $\alpha$ is the fine structure constant.
Integrating over a closed space-time with periodic boundary
conditions, we obtain the quantization
$ S_{\theta} = \int d^4x \; {\cal L}_\theta = \theta n$,
where $n$ is an integer.
Namely, $S_{\theta}$ is a topological term.
The quantization also implies that
the bulk properties depends on $\theta$ only modulo $2 \pi$.
While $S_{\theta}$ generically breaks the parity and time-reversal
symmetry, both symmetries are intact
at $\theta = 0$ and $\theta = \pi$. 

The topological term was originally
introduced in high-energy physics.
In particular, the similar term can be
defined for the quantum chromodynamics, which is
a non-Abelian gauge theory for the strong interaction.
Within the Standard Model of Particle Physics, 
there is a priori no reason to set
$\theta$ to the time-reversal invariant values.
If $\theta$ has a generic value, a strong violation
of the time-reversal or of the CP symmetry should follow,
conflicting with current experimental limitations
on the CP violation.
The solution proposed in Ref.~\cite{PecceiQuinn} to address
this problem introduces a
dynamical pseudo-scalar field, called axion field,
which couples to
$\vec{E}\cdot \vec{B}$,
so that its expectation value absorbs the parameter $\theta$.
Let us, for the moment, use the same symbol $\theta$ for the axion field.
It was shown that 
the axion field would relax into the groundstate
corresponding to $\theta =0$, thereby dynamically
recovering the time-reversal symmetry.
This solves the ``strong CP problem''.
At the same time, it leads to prediction of a new particle,
the axion, corresponding to quantum of $\theta$-field.
While the axion is a possible component of dark matter,
direct detection of the hypothetical particle so far
remains elusive.

Condensed matter physics often provides realization of
intriguing theoretical concepts, which originate in
but are rather difficult to observe experimentally
in high-energy physics. The quantum electrodynamics with 
the topological term~\eqref{eq.Ltheta} is such an example.
It was pointed out in Ref.~\cite{QHZ}
that this system at the nontrivial, time-reversal invariant angle
$\theta = \pi$ is an effective theory for topological insulators.
%The topological term represents magnetoelectric effects.
In fact, it was pointed out earlier~\cite{Hehl}
that the same theory with a smaller value of $\theta$ describes
magnetoelectric effects in Cr$_2$O$_3$~\cite{Cr2O3-ME}.
(See also Ref.~\cite{Essin}.)

The topological angle $\theta$ is a \emph{static}
parameter for a topological insulator.
Nevertheless, in Ref.~\cite{LWQZ} it was pointed out that,
when there is an antiferromagnetic order in an insulator,
the magnetic fluctuations can couple to electrons,
playing the role of the dynamical axion field.
Interesting effects due to the dynamical axion field
were predicted in the presence of an applied magnetic field.
Such a system is called ``topological magnetic insulator'' (TMI),
and it gives a condensed-matter realization of the axion
electrodynamics.

It should be noted, however, that the presence of the
dynamical axion field does \emph{not} require the
system to be a topological insulator;
a topologically \emph{trivial} insulator could have
the dynamical axion field if there is an appropriate
coupling between magnetic fluctuations and electrons.

In this paper, we study the instability of the axionic electrodynamics
in $(3+1)$ dimensions with a massive axion field, and its possible
realization in magnetic systems. In particular, we show
that the instability leads to a complete screening of an
applied electric field above a critical value,
proportional to the axion mass. 
This also leads to a spontaneous generation of a magnetic flux 
from the material.
We will discuss how this phenomenon can be detected experimentally.

\medskip
\noindent
\textit{Instability.---}
We consider the axionic electrodynamics defined by the Lagrangian density, 
in which the electromagnetic field are coupled
to the axion field $\phi$, 
\begin{equation}
\label{AxionPhoton}
\begin{split}
{\cal L} =& {\cal L}_{\rm EM}
+ \frac{\alpha}{4\pi^2} (\theta + \phi) \vec{E}\cdot \vec{B} + \cr
& + g^2 J \left( (\partial_t \phi)^2 - \nu_i^2 (\partial_i \phi)^2 - m^2 \phi^2\right) , 
\end{split}
\end{equation}
where $J$, $\nu_{i=x,y,x}$, and $m$ 
are the stiffness, velocity, and mass of the axion
% and $\alpha$ is the fine structure constant 
\cite{LWQZ}.
The time reversal symmetry is broken unless $\theta = 0$ or $\pi$.
For application to magnetic materials, the axion velocity $\nu_i$
can be anisotropic and is much smaller than the speed of light in
vacuum, which is set to unity.

Suppose we turn on the electric field $E$ in the $z$-direction and
consider fluctuations with momentum $k$ in the $x$-direction. The axion
mixes with a photon polarized in the $y$-direction, giving rise to the
dispersion,
\begin{equation}
\label{dispersion}
\begin{split}
\omega^2  = &\frac{1}{2}\left( (c'^2 + \nu^2) k^2
 + m^2 \right) \pm \cr
 &\pm \frac{1}{2} \sqrt{ \left( (c'^2-\nu^2)k^2 - m^2\right)^2 +4 m^2 c'^2 k^2 E^2/E^2_{{\rm crit}}},
\end{split}
\end{equation}
where 
\begin{equation}
\label{crit}
E_{{\rm crit}} = \frac{m}{\alpha}\sqrt{\frac{(2\pi)^3 g^2 J}{\mu}}, 
\end{equation}
$c' = 1/\sqrt{\epsilon\mu}$ is the speed of light in the medium, and $\nu = \nu_{i=2}$. In particular, 
if $E > E_{{\rm crit}}$, we find $\omega^2 < 0$ for the range of momentum,
\begin{equation}
  0 \ < \  k \ < \ \frac{m}{\nu} \sqrt{\left(\frac{E}{E_{{\rm crit}}}\right)^2 - 1} . 
\end{equation}
Namely, the electric field larger than $E_{{\rm crit}}$ is
unstable.

Consider a flat plate of the material described by (\ref{AxionPhoton}) and
sandwich it by (non-topological) insulators
with permittivity $\epsilon_0$. 
Apply a constant external electric field $E_0$ perpendicular to
the interface between the material and the ordinary insulator. 
The boundary condition at the interface gives,
\begin{equation}
\label{ebc}
\epsilon E + \frac{\alpha}{\pi}(\theta + \phi) B = \epsilon_0 E_0,
\end{equation}
where $E$ and $B$ are components of the electric and magnetic fields
normal the interface.  In addition, the conservation of magnetic flux enforces that
$B$ is continuous across the boundary.  For now, we assume
the Neumann boundary condition for the axion field $\phi$,
as it enables a simple analysis.
Later, we will study the case with the Dirichlet boundary condition,
which is relevant for physical realization.

Assuming homogeneity of the fields, the equation of motion sets,
\begin{equation}
\label{phimin}
  \phi = \frac{\alpha}{8\pi^2 g^2 J m^2} E B.
\end{equation}
The boundary condition (\ref{ebc}) then gives,
\begin{equation}
  \epsilon E = \frac{ \epsilon_0 E_0 - \alpha\theta B/\pi}
  {1 + c'^2 B^2/E_{{\rm crit}}^2}. 
\end{equation}
The energy-density ${\cal H}$ in the material can then be expressed
in terms of $B$ as,
\begin{equation}
\label{energy}
 {\cal H} = \frac{1}{8\pi \epsilon} \frac{(\epsilon_0 E_0 - \alpha\theta B/\pi)^2}{1 + c'^2 B^2/E_{{\rm crit}}^2}
 + \frac{1}{8\pi \mu} B^2. 
\end{equation}
Minimizing the energy density determines $B$. 

For example, if we turn off the axion by sending $m^2 \rightarrow \infty$, 
the minimum energy configuration is,
\begin{equation}
\label{witten}
 E = \frac{\epsilon_0}{\epsilon+ \alpha^2 \theta^2\mu/\pi^2} E_0,~~
 B = \epsilon_0 \mu \frac{\alpha\theta/\pi}{\epsilon  + \alpha^2 \theta^2\mu /\pi^2}
   E_0
\end{equation}
as expected from the Witten effect \cite{Witten}. 
We note that, although the bulk theory defined by eq.~\eqref{AxionPhoton}
has $2\pi$-periodicity in $\theta$,
the results here are no longer periodic in $\theta$
since the boundary breaks the periodicity.

\begin{figure}[htbp]
\begin{center}
\includegraphics[width=7cm]{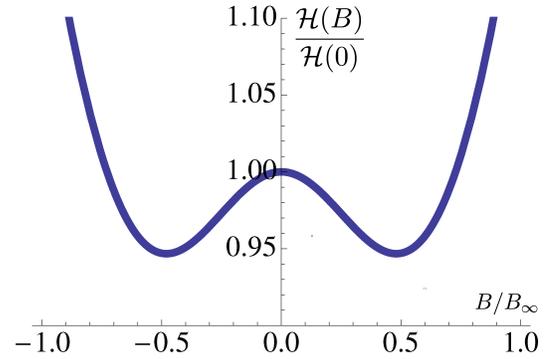}
\caption{The
 energy density~\eqref{energy} for $\theta=0$ and 
$\epsilon_0 E_0 = 1.3 \epsilon  E_{\rm crit}$,
which is slightly above the critical value 
$\epsilon_0 E_0= \epsilon E_{\rm crit}$. 
The energy ${\cal{H}}(B)$ shows a double-well structure 
and the spontaneous breaking of the time reversal symmetry 
$B \rightarrow -B, \phi \rightarrow - \phi$.
The magnetic field corresponding
to the potential minima is spontaneously generated, resulting
in the screening of the electric field to $E = E_{\rm crit}$.
The magnetic field is normalized by the asymptotic value
$B_\infty$ given in eq.~\eqref{asymptotic}.}
\label{fig.SSB}
\end{center}
\end{figure}

Let us first 
analyze the energy density (\ref{energy}) when 
$\theta = 0$. When the applied electric field is lower than 
the critical value as in  
$\epsilon_0 E_0 < \epsilon E_{{\rm crit}}$,
the only solution is,
\begin{equation}
\label{trivial}
E=\frac{\epsilon_0}{\epsilon}E_0, ~~B =0.
\end{equation}
However, if the external electric field $E_0$ is raised above $\frac{\epsilon}{\epsilon_0} E_{{\rm crit}}$,
we find states with lower energy given by,
\begin{equation}
\label{screened}
 E = E_{{\rm crit}}, ~~
  B = \pm \sqrt{\mu E_{{\rm crit}} (\epsilon_0 E_0 - \epsilon E_{{\rm crit}})}. 
\end{equation}
Thus, there is a second order phase transition at $\epsilon_0 E_0 =
\epsilon E_{{\rm crit}}$.  Above this value, the electric field inside
the material stays constant and the magnetic field is increased instead.
We show an example of the energy density as a function of $B$ above the
critical value, in Fig.~\ref{fig.SSB}. 

Now, let us turn to the case with $\theta \neq 0$. When $\epsilon_0 E_0
\ll \epsilon E_{{\rm crit}}$, there is a unique minimum energy
configuration, which reduces to (\ref{witten}) in the limit of $E_0
\rightarrow 0$.  For $\epsilon_0 E_0 \gg \epsilon E_{{\rm crit}}$, the
energy density has two local minima in $B$.  Since the time-reversal
symmetry is broken explicitly for $\theta \neq 0$ ($\theta = \pi$ does
not preserve the time reversal-symmetry in the presence of the
boundary), the two minima have different energies.
%as shown in Fig.~\ref{fig.noSSB}.
In the limit of $E_0 \rightarrow \infty$, the more stable of the two behaves as
\begin{equation} 
\label{asymptotic}
  E \rightarrow E_{{\rm crit}}, ~~
  B \rightarrow B_{\infty} = \sqrt{ \epsilon_0 \mu E_0 E_{{\rm crit}}}.
\end{equation}
One can show that the
configuration (\ref{witten}) for small $E_0$ is smoothly connected
to (\ref{asymptotic}) for large $E_0$. Namely, the phase transition at  
$\epsilon_0 E_0 = \epsilon E_{{\rm crit}}$  is smoothened out 
when $\theta \neq 0$. 
However, for realistic values of $\theta \sim \pi$,
the smoothing effect is small and the solution is 
similar to the one at $\theta=0$ with a second-order phase transition, except that
the groundstate is chosen uniquely for any $E_0$.
%In Fig.~\ref{fig.noSSB}, $\theta = 20 \pi$ is chosen in order
%to make the effect visible.

\begin{figure}[htbp]
\begin{center}
\includegraphics[width=7cm]{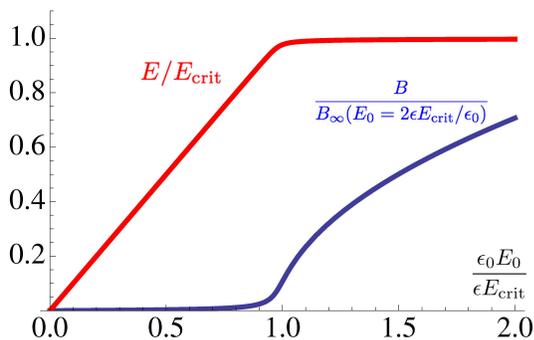}
\caption{The electric field
$E$ (red) inside the material and 
the induced magnetic field $B$ (blue) as functions of
the external field $E_0$, evaluated at $\theta=\pi$.
$B$ and $E$ are determined by the global minimum of the energy
density~\eqref{energy}.
The fields $E_0, E$ and $B$ are normalized by
$\frac{\epsilon}{\epsilon_0} E_{\rm crit} $, $E_{\rm crit}$, and
$B_\infty(E_0 = 2 \frac{\epsilon}{\epsilon_0} E_{\rm crit})$, respectively.
The behavior appears to indicate a second-order phase
transition, although the transition is smoothed out
by the effect of non-zero $\theta$.}
\label{fig.E0_vs_B_Pi}
\end{center}
\end{figure}

We note that, in realistic systems, higher order
terms of $\phi$ are expected in the effective theory,
since the range of ordered magnetic moment,
which corresponds to shift of $\phi$, is bounded.
However, although it will be important for $E \gg E_{\rm crit}$,
this does not affect the existence of the transition,
because $\phi , B \sim 0$ in the vicinity of the transition.

For experimental realization of the screening effect discussed below,
the Dirichlet boundary condition $\phi = \phi_0$
will also turn out to be relevant.
In this case, the solution is no longer
uniform in the $z$-direction.
Nevertheless, away from the boundary,
the solution asymptotically approaches to
the stationary configuration, which is given by 
the solution obtained in the above for the Neumann
boundary condition, with the replacement $\theta \rightarrow -\phi_0$.
The screening of the electric field occurs in the transient
region with a finite length.

If we ignore the coupling to the electromagnetic field,
the axion field $\phi$ approaches
to the stationary solution exponentially,
with the decay length of $\nu/m$.
However, we should take into account mixing of the axion and
the photon via the coupling $\phi E B$.
This gives the axion field the effective mass $m_{\rm eff}$, which
is given by
\begin{equation}
m_{{\rm eff}}^2 = m^2 + \frac{\alpha^2 B^2}{8\pi^3 g^2 J
\epsilon}.
\end{equation}
When $\phi_0 = - \theta = 0$ and the applied electric field is
above the critical value $\epsilon_0 E_0 > \epsilon E_{{\rm crit}}$, the
magnetic field is given by (\ref{screened}). When $\phi_0 \neq 0$, this
is the asymptotic value for $E_0 \rightarrow \infty$. 
Substituting this value of $B$ in the above, we find that the
screening occurs within the lengthscale 
\begin{equation} 
  \frac{\nu}{m_{{\rm eff}}} = \frac{\nu}{m}
\sqrt{\frac{\epsilon E_{{\rm  crit}}}{\epsilon_0 E_0}}.
\label{eq.screening_length}
\end{equation}
Physically, the screening of the electric field occurs
because of the induced charge density
$\propto \vec{\nabla}\phi \cdot \vec{B}$
in the axion electrodynamics \cite{Wilczek}.
Namely, the axion field $\phi$ is shifted inside the material
creating the gradient $\vec{\nabla} \phi$ near the boundary.
By generating the magnetic field $\vec{B}$, this induces
a charge density at the boundary, screening the electric
field inside the material.

From this physical picture based on screening,
we can also understand why the second-order transition
occurs when $\phi_0 = 0$.
Under a given applied electric field, the sign
of the charge needed to screen the electric field
is uniquely determined.
However, induced charge is proportional to
$\partial_z \phi\cdot B$, which is invariant under
the time reversal
$\phi \rightarrow - \phi, B \rightarrow -B$.
The symmetry is preserved for the boundary condition
$\phi_0=0$, and there is no preferred sign of $B$.
The system breaks the symmetry spontaneously to
produce screening charge, for $E>E_{\rm crit}$.

On the other hand, $\phi_0 \neq 0$ introduces a gradient
of $\phi$ near the boundary, choosing the
preferred sign of $B$ to produce screening charge.
Thus the symmetry is broken explicitly and
the phase transition is smeared.

\medskip
\noindent
\textit{Possible Experimental Realization in Magnetic Materials.---} 
Let us discuss possible realization of the
axionic instability in condensed matter systems.
In Ref.~\cite{LWQZ}, Bi$_2$Se$_3$ doped with $3d$
transition metal elements such as Fe
(Bi$_2$Se$_3$-Fe hereafter) is discussed
as a candidate for TMI, which is described 
by the axionic electrodynamics.
Although the mechanism proposed in this paper is not
restricted to any particular system,
we shall examine possible physical realization
using \BiSe -Fe as a reference.
Because of the magnetic doping,
a magnetic order $M^-$, which is ferromagnetic in
the $xy$-plane and antiferromagnetic along the $z$-direction,
may appear~\cite{LWQZ}.
In the following, we assume that
the electric field is applied along the $z$-axis.

The (relative) permittivity, axion mass, and
axion coupling in \BiSe -Fe
are estimated~\cite{LWQZ} to be
$\epsilon \sim 100$,
$m \sim 2 \; \mbox{meV}$, and
$(2\pi)^3 g^2 J = \alpha ( 0.4 \; \mbox{T/meV})^2$.
This yields a rather high value of
$E_{\rm crit} = 2.4 \times 10^8$ $\mbox{V/m}$,
which is above the breakdown field of typical semiconductors.

The critical field $E_{\rm crit}$ is reduced for smaller
axion mass $m$. In the ordered phase,
the axion mass is proportional to the spontaneous
magnetic order $M^-$~\cite{LWQZ}.
Thus, the axion mass may be reduced by tuning the system
close to the critical point, so that the antiferromagnetic
order becomes small.
If, for example, $E_{\rm crit} \sim 10^7 \; \mbox{V/m}$ and a TMI film
of thickness $\sim 10^{-8} \; \mbox{m}$ is used, the
voltage difference across the system is of the order of $0.1 \; \mbox{V}$.
The corresponding energy is well
below the band-gap $\sim 0.3 \; \mbox{eV}$ of \BiSe ,
justifying the low-energy description.

However, TMIs such as \BiSe -Fe are expected to
have surface electronic states.
For the undoped topological insulator (such as \BiSe ),
the surface states are described by massless Dirac fermions.  
In the doped TMI, the surface state has a gap $m_5$
due to the time-reversal symmetry breaking.
In Bi$_2$Se$_3$-Fe, it was estimated~\cite{LWQZ} that
$m_5 =1$ meV and $m=2$ meV; these are of the same
order of the magnitude.
However, in order to suppress the screening effect
by the surface states and to enhance the effect by the
dynamical axion in the bulk, 
it is desirable to keep the surface Dirac mass $m_5$ large
while tuning the system close to the criticality
to reduce the axion mass $m$.
This could be achieved by sandwiching the TMI by ferromagnets,
which enforces the magnetic order near the
surfaces~\cite{QHZ} and determines the boundary
value $\phi_0$.

The axion mass $m$ and the spin-wave velocity
$\nu$ are expected to be of the order of
$U$ and $J_{ex} a$,
respectively, where $J_{ex}$ is the effective exchange
interaction and $a$ is the lattice constant.
Thus, the screening length~\eqref{eq.screening_length}
is expected to be of the order of the lattice constant
when $U$ is sufficiently large.
This implies that the present effect could be observed
in thin film samples.

Experimentally, the instability discussed in this paper 
is observed as an increase of capacitance
above the critical electric field.
Another experimentally observable consequence is the
generation of magnetic field as in eq.~\eqref{asymptotic}.

In the above scenario, 
the necessity to gap out the surface Dirac mode
by sandwiching the TMI by ferromagnets may pose
an additional complication. 
If we begin with a topologically \emph{trivial}
insulator, there is no surface Dirac mode.
In fact, for the instability discussed here, it is not essential
that the system is based on a nontrivial topological insulator.
Even in a topologically trivial insulator,
if there is an appropriate coupling with the
magnetic order and the electrons, the magnetic fluctuations
can play the role of the axion field,
following the argument parallel to that of Ref.~\cite{LWQZ}.
The difference is that, starting from a trivial insulator
we expect $\theta \sim 0$, instead of $\theta \sim \pi$ in a TMI.
However, as long as the dynamical axion $\phi$ is present
in the low-energy effective theory~\eqref{AxionPhoton},
the mechanism proposed in this paper should work.
The class of magnetic materials with a dynamical
axion field may be called axionic insulators.

Cr$_2$O$_3$ is a topologically trivial insulator with
the band gap of 3.4 eV, and a magnetic long-range order.
As we have mentioned, it exhibits magneto-electric
effects, which can be described by the axion electrodynamics.
It is also suggested that the antiferromagnetic
fluctuations in Cr$_2$O$_3$
play the role of the dynamical axion field~\cite{Qi-private}.
The present mechanism could in principle be realized in
such a system.
The large band gap itself is advantageous for observation
of the effect, as the material can withstand stronger
electric field. However, the large band gap also implies
a large coupling $g$, which is approximately
proportional to the band gap.
This results in a larger critical field because of
eq.~\eqref{crit}, cancelling the advantage.

The axion mass $m$ is proportional to the
on-site Coulomb repulsion $U$ and the magnetic order~\cite{LWQZ}.
The Coulomb repulsion $U$ in Cr$_2$O$_3$ is estimated to be
about 5 eV~\cite{Cr2O3-U};
thus the axion mass $m$ is expected to be of the same order,
resulting in a too large value of $E_{\rm crit}$.
In order to observe the instability, we need to find a
system with smaller axion mass. Again, this could be achieved
by tuning the system near the magnetic criticality,
for example by mixing with nonmagnetic ions.

We would like to end with a remark on earlier papers,
where similar instabilities were discussed. 
It was pointed out in Ref. \cite{NOP,OP} that the 
Maxwell theory with the Chern-Simons term in $(4+1)$ 
space-time dimensions is unstable in the presence of
a constant electric field. Subsequently, a similar instability was
also found in its dimensional reduction to $(3+1)$ dimensions: 
the massless axion electrodynamics \cite{DG,BJLL}. 
(See also the earlier work \cite{arXiv:0704.1604}.)
In this paper, we extended these results for the
$(3+1)$ dimensional theory with a massive axion field,
clarified the eventual fate of the unstable gauge theory
in the Minkowski space, and proposed
its possible experimental realization. Although 
these earlier papers were in the context of the 
AdS/CFT correspondence, nothing in this paper 
assumes the AdS/CFT correspondence or 
depends on results obtained by it.

%%%%%%%%%%%%%%%%%%%%%%%%%%%%%%%%%%%%%%%%%%

\medskip
We thank Y.~Ando, O.~Bergman, J.~Gauntlett, D.~Hsieh, J.~Maldacena, and X.-L.~Qi
for useful discussion. We thank the hospitality of 
the Aspen Center for Physics (NSF grant 1066293) and Kavli Institute for Theoretical
Physics, UC Santa Barbara (NSF grant PHY05-51164). The work of HO is supported in part 
by U.S. DOE grant DE-FG03-92-ER40701, the 
WPI Initiative of MEXT of Japan, and JSPS Grant-in-Aid for Scientific 
Research C-23540285.  
The work of MO is supported in part by
JSPS Grant-in-Aid for Scientific Research No. 21540381.

%%%%%%%%%%%%%%%%%%%%%%%%%%%%%%%%%%%%%%%%%%%%%%%%%%%%%%%%%%%%%%%%%%%%%%%%%%%%%%%%%%

\end{document}